\documentclass[
  journal=nws,
  manuscript=research-article,
  year=2026,
  volume=1,
]{cup-journal}

\usepackage{amsmath}
\usepackage[nopatch]{microtype}
\usepackage{xcolor}
\usepackage{array}
\usepackage{booktabs}

\usepackage{graphicx}
\usepackage{hyperref}

\title{Scaling laws in empirical networks}

\author{Upasana Dutta}
\affiliation{Department of Computer and Information Science, University of Pennsylvania, Philadelphia, United States}
\email[Upasana Dutta]{upasanad@seas.upenn.edu}

\author{Alexander Ray}
\affiliation{Department of Computer Science, University of Colorado Boulder, Boulder, United States}

\author{Aaron Clauset}
\affiliation{Department of Computer Science, University of Colorado Boulder, Boulder, United States}
\alsoaffiliation{BioFrontiers Institute, University of Colorado, Boulder, United States}
\alsoaffiliation{Santa Fe Institute, Santa Fe, United States}


\makeatletter
\def\ps@firstpage{%
  \let\@oddhead\@empty
  \let\@evenhead\@empty
  \def\@oddfoot{\hfil\thepage\hfil}%
  \let\@evenfoot\@oddfoot
}
\makeatother
\begin{document}

\begin{abstract}
How does the shape of a network change as its size increases? Although random graph models provide some expectations for such ``scaling behaviors'' in the structure of networks, relatively little is known about how empirical network structure scales with network size or how well random graphs explain those empirical patterns. Using a large, structurally diverse corpus of networks from four scientific domains, we first characterize the empirical scaling laws of real-world networks, considering how mean degree, transitivity, mean geodesic distance, and degree assortativity vary with network size. We show that networks from all four scientific domains exhibit a consistent set of scaling laws on these measures of network structure, but with differing scaling rates. We then assess the extent to which these empirical scaling laws are explained by three random graph models with different structural assumptions, showing that configuration model random graphs are a remarkably good model of network scaling behavior, although null models with modular structure are slightly better. These findings identify a new set of common patterns in the network structure of complex systems, provide new validation targets for models of network structure, and shed new light on the role of randomness in shaping the large-scale structure of networks.
\end{abstract}

\section{Introduction}
Networks are commonly used to represent pairwise and other interactions among parts of complex systems in a wide variety of scientific domains, including friendships between people in social networks, binding affinity among proteins in biological networks, and hyperlinks between web pages in informational networks. Characterizing the statistical patterns among a network’s connections—sometimes called ``exploratory network analysis”— provides useful insights about the shape of a particular network, e.g., how densely connected it is, how quickly information might be able to spread across it, and how clustered or modular it is, etc. These statistical descriptions of network structure are universally used to describe, explain, and make predictions about real-world networks, and are the primary way we test structural hypotheses using network null models (\cite{hobson2021guide, fosdick2018configuring, dutta2025sampling}) or evaluate the results of mathematical or computational models of network structure (\cite{goldenberg2010survey, rottjers2021null}).

In practice, however, the values of network summary statistics like mean degree, mean geodesic distance, transitivity, and degree assortativity tend to vary widely across networks, reflecting broad structural diversity both within and across scientific domains (\cite{newman2010networks, dorogovtsev2022nature}). Beyond a handful of stylized notions about what values we should expect for these measures on different types of network, relatively little is known quantitatively about how they vary across network scales or domains, and whether that variation is systematic. For instance, conventional wisdom tells us that social networks have higher transitivity (clustering) coefficients and stronger modular structure than non-social networks (\cite{newman2003social}), that many real-world networks have ``short'' mean geodesic distances (\cite{watts1998collective, newman2010networks}), that social networks tend to be assortative by degree while biological networks tend to be disassortative by degree (\cite{newman2002assortative}). While each measure has clear interpretations and effects on the dynamics, behavior, and efficiency (\cite{latora2001efficient}) of a system, a better understanding of their systematic variation across types and sizes of networks would shed new light on how generative processes influence the large-scale structure of complex systems and the extent to which fundamental laws underpin the structure of real-world networks. 

Here, we investigate a set of simple questions about the ``scaling behavior'' of network structure, whose answers would have broad implications for network analysis and modeling efforts. How does the shape of a network change as its size increases? For instance, do networks get sparser or denser with size? How quickly do the lengths of geodesic paths grow? Does the tendency to connect to nodes of similar degree increase as networks increase in size? And, do these patterns depend on whether the network in question is a social network, a biological network, or a technological network? Our approach to answering these questions is to characterize the ``scaling laws" of network structure, i.e., to consider how a measure of network structure varies systematically with network size. This approach has been highly successful in identifying important patterns in other scientific domains, such as Kleiber’s Law for how an organism’s metabolic rate scales up with its body mass (\cite{kleiber1932body, dodds2001re, delong2010shifts}), and Zipf’s Law and related patterns for how social and economic indicators scale up with city population (\cite{bettencourt2007growth, keuschnigg2019urban}). Such empirical scaling laws have proved highly generative for developing new macro-level theoretical models and for using deviations from the laws to identify as-yet-unexplained empirical phenomena.

The use of scaling laws in network science requires a large number of real-world networks, drawn from a wide variety of sources. For much of the history of network science, the availability of such corpora was limited. Here, we use a structurally diverse corpus of 254 networks, drawn from the list of research-quality network data sets in the Index of Complex Networks (ICON; \cite{clauset2016colorado}). This corpus includes networks from four scientific domains where ICON has good coverage: social networks, biological networks, informational networks, and technological networks, ranging in size from about $10^1$ nodes to about $10^7$ nodes. The diversity and size of this corpus allows us to quantitatively characterize the scaling behaviors of empirical networks and extract generalizable insights about a diverse class of networks. Details on the inclusion criteria for this corpus are given in Section~\ref{sec:network_scaling}.

Historically, random graph models like the Erdős–Rényi model (\cite{erdds1959random, erdHos1960evolution}) and the configuration model (\cite{newman2001random}) have defined our expectations for how different aspects of network structure should change as a network scales up in size. For instance, mathematical analyses of the Erdős–Rényi random graph model, which is a random graph conditioned on a specified mean degree, predict that the mean geodesic distance, or average length of a shortest path, should scale like $O(\log n)$ while the transitivity (clustering coefficient) should scale like $O(1/n)$ (\cite{newman2010networks}). The configuration model, which is a random graph conditioned on a specified degree sequence (and is thus a richer model of structure), makes predictions with similar functional forms, but with scaling rates that depend on the mean and variance of the selected degree sequence (\cite{newman2010networks}). When compared to specific real-world networks, random graphs with fixed mean degree or degree sequences often fail to reproduce the high levels of transitivity (clustering coefficient) that are the hallmark of social networks (\cite{newman2003social, newman2001random}), even as they succeed in reproducing the ubiquitous pattern of short geodesic (shortest path) distances that appear in networks from nearly all scientific domains. This mismatch is, in part, what led to the development of the ``small-world” network model (\cite{watts1998collective}), which showed that high transitivity and short geodesic distances can co-occur in a network, as is seen in many real-world networks (\cite{strogatz2001exploring, bassett2006small, stam2007small, latora2002boston, wagner2001small, kogut2001small, adamic1999small}).

However, it is not known how well real-world networks match the random graph predicted scaling behaviors, and thus, it is unclear how well random graphs can serve as reference models in exploratory network analyses. For instance, if the configuration model’s predicted scaling behavior for transitivity largely matches what we see in empirical data, we may conclude that at a macro level, the density of triangles in large-scale networks can be attributed to degree-corrected randomness in connectivity at the network scale. Alternatively, if the empirical scaling law exhibits the same functional form but a different scaling rate than the configuration model, we may conclude that some additional structural mechanism biases the randomness at the macro level in these networks.

First, we characterize the empirical scaling laws of a network’s mean degree, mean geodesic distance, transitivity (global clustering coefficient), and degree assortativity in each of four scientific domains, showing that networks in different domains follow a consistent set of empirical scaling laws, but with somewhat different scaling rates that reveal interesting differences in the structure of networks by domain. We then compare these empirical scaling laws to theoretical scaling laws predicted by three random graph models with increasing complexity: fixed mean degree, fixed degree structure, and fixed degree structure with modules (communities). This comparison shows that the empirical scaling laws in different scientific domains deviate systematically from those predicted by the simplest random graph model, while being captured fairly well in many instances by the configuration model and the modular random graph models. These findings demonstrate how empirical scaling laws provide a useful test for validating network models, and show that some random graph models represent remarkably good models of the large-scale structure of networks. 

\section{Network Data, Scaling Laws, and Random Graph Models}
\label{sec:network_scaling}

The corpus we use includes 254 networks from ICON (\cite{clauset2016colorado}). These networks range in size from $10^1$ to $10^7$ nodes (Table I), and are drawn from the social, biological, informational, and technological domains. In social networks, nodes are typically people, and edges represent some kind of social interaction, e.g., offline friendship, online friendship, etc. Biological networks are a diverse category in which nodes might be genes, metabolites, proteins, neurons, or even entire species, and edges represent regulatory actions among genes, involvement in an enzyme-catalyzed reaction, binding among proteins, synaptic connections among neurons, or a transfer of nutrients by predation in a food web, respectively. Informational networks are a more abstract category in which nodes can be documents, words, concepts, etc., and edges might represent citations, co-occurrence, or relatedness. Finally, technological networks are typically some form of computer network, in which nodes can be computers, servers, internet addresses, mobile phones, etc., and edges indicate a direct transfer of information. This variety of networks from diverse scientific domains provides a substantive sample of the underlying distribution of all empirical networks from which to draw general conclusions. Although the underlying networks included in the corpus may not be simple, all networks analyzed here were simplified by ignoring edge direction and weight, collapsing multiedges, and removing self-loops. See~\ref{sec:appendix_materials} for detailed inclusion criteria and simplification procedure.

We use scaling laws to estimate the functional form and growth rates of the asymptotic behavior of network summary statistics. Comparing the forms and estimated parameters of these functions across scientific domains and between different random graph models provides a quantitative way to distinguish qualitatively different kinds of behavior in how the shape of networks change with size. In particular, we consider two forms of scaling functions: (1) a power law form $y = a \times n^b$, in which the network measure $y$ varies as a power law in the number of nodes $n$, with $b$ denoting the ``scaling coefficient” and $a$ denoting the ``scaling intercept”; and (2) a logarithmic form $y = a + b \times \log_{10} n$, in which the network measure grows logarithmically with the number of nodes $n$, $b$ denoting the ``scaling coefficient” and $a$ denoting the ``scaling intercept.” Visually, the power law form is a straight line on a log-log plot, while the logarithmic form is a straight line on a semi-log plot.


\begin{table}[t!]
\centering
\begin{tabular}{lccccc}
\toprule
\textbf{Domain} & \textbf{Simple} & \textbf{Weighted} & \textbf{Directed} & \textbf{Multi} & \textbf{All} \\
\midrule
\textcolor{orange}{Social} & 46 & 2 & 32 & 21 & 88 \\
\textcolor{blue}{Biological} & 52 & 3 & 25 & 16 & 88 \\
\textcolor{green!50!black}{Informational} & 3 & 0 & 29 & 16 & 37 \\
\textcolor{violet}{Technological} & 6 & 1 & 32 & 15 & 46 \\
\midrule
\textbf{All} & \textbf{107} & \textbf{7} & \textbf{118} & \textbf{68} & \textbf{254} \\
\bottomrule
\end{tabular}
\caption{Summary of the 254 networks that comprise our structurally diverse corpus, which spans four distinct scientific domains. Simple networks are defined as unweighted, undirected graphs without multi-edges.}
\label{tab:table1}
\end{table}

In the case of power law scaling, when the scaling parameter $b<0$, the network measure decays systematically toward 0 as network size scales up, while if $b>0$, the measure grows asymptotically. For instance, in the Erdős–Rényi random graph model, the global clustering coefficient (transitivity) exhibits a scaling law like $C = C_0 \times n^{-1}$, indicating that as networks get larger, their density of triangles steadily decreases (\cite{newman2010networks}). In the case of logarithmic scaling, when $b>0$, the measure is increasing, albeit very slowly (logarithmically) with network size. 

We estimate the scaling behavior of a set of networks using four standard measures of network structure: (1) the mean degree $\langle k \rangle$, (2) the mean geodesic distance $\langle \ell \rangle$, (3) the global clustering coefficient $C$, and (4) the degree assortativity $r$. Mathematical details for the calculation of each network summary statistic are provided in \ref{sec:appendix_network_measures}. These network measures characterize the shape and connective properties of networks in the corpus, allowing us to draw conclusions about how that shape systematically varies with network size and scientific domain. Scaling law parameters for each of the parametric forms are estimated via maximum likelihood using OLS regression.

Finally, we compare the estimated empirical scaling laws with random graph scaling laws for three choices of random graph ``null models.'' This comparison allows us to quantitatively assess the degree to which the observed empirical scaling behavior can be explained by the assumptions underlying each random graph model. The three null models are (1) the $G(n,m)$ model of Erdős–Rényi random graphs (\cite{erdHos1960evolution}), in which a fixed number of edges $m$ are distributed uniformly at random among $n$ nodes; (2) the configuration model (\cite{fosdick2018configuring, dutta2025sampling}), which is a random graph conditioned on a fixed degree sequence $\vec{k}$ over $n$ nodes; and (3) a variation of the degree-corrected stochastic block model (DC-SBM) (\cite{karrer2011stochastic}), which fixes the degree sequence and the modular (community) structure of the network (\cite{peixoto2017nonparametric}). \ref{sec:appendix_null_models} provides technical details for the estimation and generation of each random graph model. We note that each of these models fixes the number of edges in a network; in \ref{sec:appendix_robustness} we show results for random graph models that match an empirical network’s characteristics only in expectation.

\begin{figure*}[t!]  
    \centering
    \includegraphics[width=0.99\textwidth]{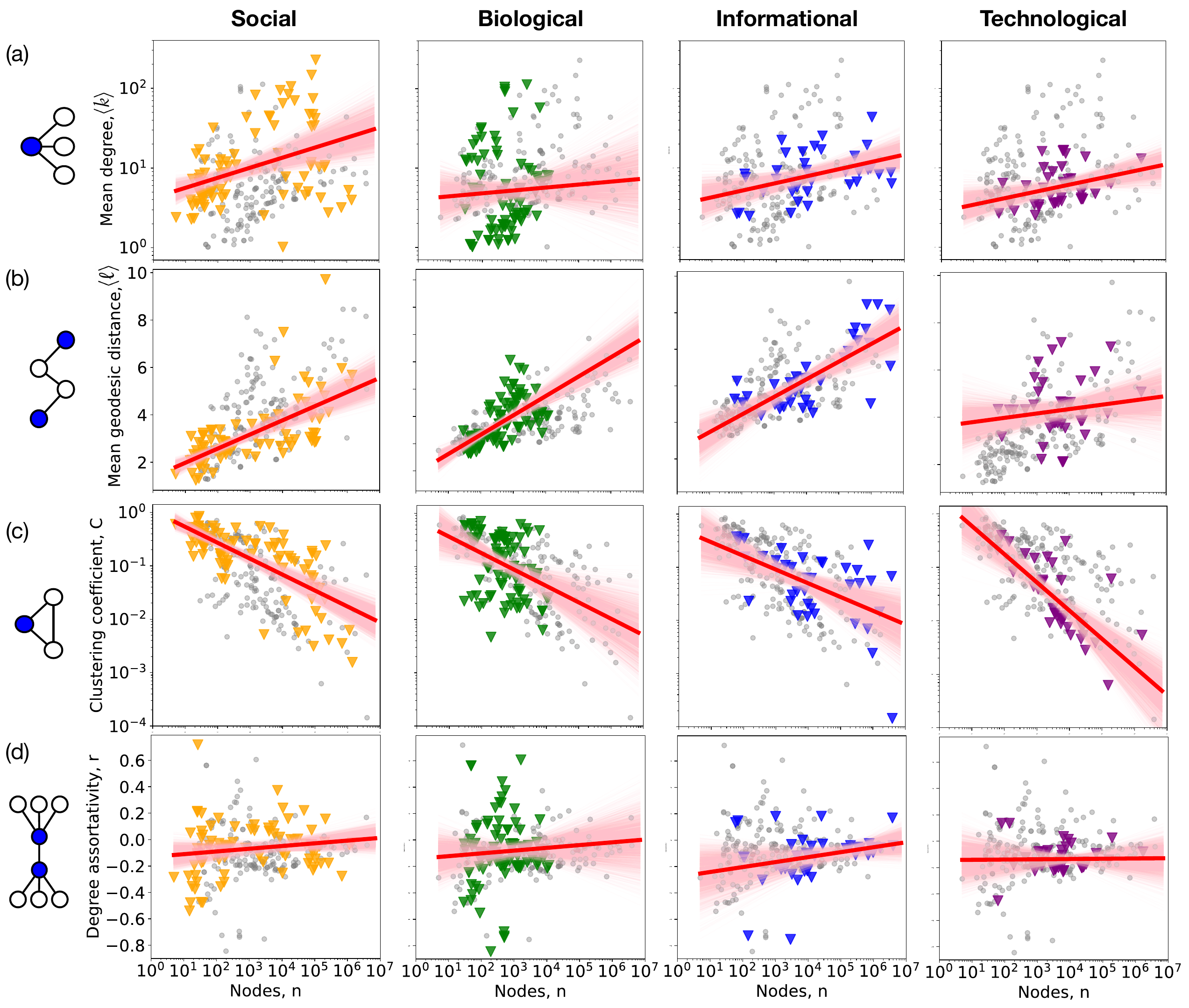}
    \caption{(a) mean degree, (b) mean geodesic path length, (c) clustering coefficient, and (d) degree assortativity of the 254 empirical networks in our corpus as a function of the number of nodes, across the social, biological, informational, and technological domains. The colored triangles represent networks of the particular domain, while the gray circles in the background represent the remaining networks outside each domain.}
    \label{fig:empirical_scaling}
\end{figure*}

For each empirical network $G_i$, we compute the expected values for three out of the four network measures (mean geodesic distance $\langle \ell \rangle$, global clustering coefficient $C$, and degree assortativity $r$) under a particular null model when parameterized by $G_i$. We do not compute the expected values for the mean degree $\langle k \rangle$ since all three null models fix the mean degree of the network they generated. In particular, we fit the null model to $G_i$, e.g., in the $G(n,m)$ random graph model, we set its parameters $n$ and $m$ equal to those of $G_i$, generate 50 synthetic networks, and then compute the mean values for each of the three network measures. In our numerical experiments, we find that using more than 50 synthetic networks does not result in qualitatively different results. Repeating this process for every network in the corpus, and for each null model, produces three sets of theoretical $y$ values. We then estimate the corresponding scaling law parameters $a,b$ for that network measure to obtain the null model’s expected scaling behavior. The closer the parameters of a null model’s scaling behavior are to the empirically estimated parameters, the better that model explains the observed scaling behavior. 

\subsection{Empirical Scaling Laws} 
We present scaling behaviors of networks across scientific domains as an empirical study of how summary statistics of real-world networks change with network size.

\paragraph{Average Degree} 
All of the networks that compose our corpus are relatively sparse. Among these networks, we observe that the mean degree $\langle k \rangle$ appears to increase slowly relative to network size $n$, in social, biological, informational, and technological networks, as shown in Figure~\ref{fig:empirical_scaling}a. This positive correlation between mean degree and network size is similar to the observations in Ref.~\cite{broido2019scale}, using a similarly diverse corpus of empirical networks, and stands in contrast to the typical random graph assumption that mean degree is asymptotically constant, i.e., independent of network size (\cite{jackson2008social, newman2010networks, eric2009statistical, menczer2020first}).

When fitted with the power law scaling function (Table~\ref{tab:table2}), we find that social networks exhibit the fastest scaling behavior in network density, both with a higher intercept and a substantially higher scaling parameter ($b=0.13\pm0.04$ vs. the next highest parameter $b=0.09\pm0.03$ for informational networks). That is, social networks tend to be denser, on average, and exhibit a stronger pattern of densification as they scale up. The weakest densification pattern is in biological networks ($b=0.04\pm0.07$). However, we note that these scaling relationships are simply general patterns and there is substantial variation around the fitted scaling laws in every domain. For instance, in informational and technological networks, the observed mean degrees can vary by almost a factor of 10 across networks of a given size $n$. In social and biological networks, that variation can be nearly a factor of 100, indicating broad variation around the central tendency of the estimated scaling laws. The upper end of the mean degree distribution in social networks appears to be driven by high-degree online systems, where connections are cheap compared to those in offline systems.

\begin{table}[t!]
    \centering
    \LARGE
    \resizebox{0.99\linewidth}{!}{ 
    \begin{tabular}
    {|c|c|c|c|c|}
        \toprule
        & \textbf{Mean degree} & \textbf{Mean geodesic distance} & \textbf{Clustering coefficient} & \textbf{Degree assortativity} \\
        & $\langle k \rangle = \textbf{\textcolor{red}{a}} \times n^\textbf{\textcolor{cyan}{b}}$ & $\langle \ell \rangle = \textbf{\textcolor{red}{a}} + \textbf{\textcolor{cyan}{b}} \times \log_{10} n$ & $C = \textbf{\textcolor{red}{a}} \times n^{\textbf{\textcolor{cyan}{b}}}$ & $r = \textbf{\textcolor{red}{a}} + \textbf{\textcolor{cyan}{b}} \times \log_{10} n$ \\
        \midrule
        \textcolor{orange}{\textbf{Social}} 
        & 
        $a = \textbf{\textcolor{red}{4.2}}(0.84), \, b = \textbf{\textcolor{cyan}{0.13}}(0.04)$ 
        & 
        $a = \textbf{\textcolor{red}{1.38}}(0.19), \, b = \textbf{\textcolor{cyan}{0.60}}(0.08)$ 
        &
        $a = \textbf{\textcolor{red}{1.07}}(0.2), \, b = \textbf{\textcolor{cyan}{-0.3}}(0.03)$ 
        & 
        $a = \textbf{\textcolor{red}{-0.13}}(0.05), \, b = \textbf{\textcolor{cyan}{0.02}}(0.01)$ \\
        
        \textcolor{green!45!black}{\textbf{Biological}} 
        & 
        $a = \textbf{\textcolor{red}{4}}(2.4), \, b = \textbf{\textcolor{cyan}{0.04}}(0.07)$ 
        & 
        $a = \textbf{\textcolor{red}{-0.18}}(0.43), \, b = \textbf{\textcolor{cyan}{1.42}}(0.18)$ 
        & 
        $a = \textbf{\textcolor{red}{0.7}}(0.5), \, b = \textbf{\textcolor{cyan}{-0.3}}(0.1)$ 
        & 
        $a = \textbf{\textcolor{red}{-0.14}}(0.1), \, b = \textbf{\textcolor{cyan}{0.02}}(0.04)$ \\

        \textcolor{blue}{\textbf{Informational}} 
        & 
        $a = \textbf{\textcolor{red}{3.5}}(1.2), \, b = \textbf{\textcolor{cyan}{0.09}}(0.03)$ 
        & 
        $a = \textbf{\textcolor{red}{0.5}}(0.68), \, b = \textbf{\textcolor{cyan}{0.96}}(0.18)$ 
        & 
        $a = \textbf{\textcolor{red}{0.53}}(0.6), \, b = \textbf{\textcolor{cyan}{-0.3}}(0.09)$ 
        & 
        $a = \textbf{\textcolor{red}{-0.3}}(0.1), \, b = \textbf{\textcolor{cyan}{0.04}}(0.03)$ \\

        \textcolor{violet}{\textbf{Technological}} 
        & 
        $a = \textbf{\textcolor{red}{2.8}}(0.9), \, b = \textbf{\textcolor{cyan}{0.08}}(0.03)$ 
        & 
        $a = \textbf{\textcolor{red}{3.6}}(1.71), \, b = \textbf{\textcolor{cyan}{0.18}}(0.19)$ 
        & 
        $a = \textbf{\textcolor{red}{2}}(3), \, b = \textbf{\textcolor{cyan}{-0.5}}(0.1)$ 
        & 
        $a = \textbf{\textcolor{red}{-0.14}}(0.1), \, b = \textbf{\textcolor{cyan}{0.002}}(0.03)$ \\
        \bottomrule
    \end{tabular}
    } 
\caption{Scaling behavior of mean degree $\langle k \rangle$, mean geodesic distance $\langle \ell \rangle$, clustering coefficient C, and degree assortativity r, as a function of the number of nodes n for 254 empirical networks. Scaling behaviors are listed for the social, biological, technological, and informational network domains. Uncertainties correspond to standard deviations of 10000 bootstrap samples of the fitted coefficients.}
\label{tab:table2}
\end{table}


\paragraph{Mean Geodesic distance} 
In all four scientific domains, the mean geodesic distance $\langle\ell\rangle$ of a network appears to scale logarithmically with the size of the network $n$ (Figure~\ref{fig:empirical_scaling}b), implying that all of these networks are highly ``compact,” i.e., the typical distance that any information or signal must travel to get from one part of the network to another is very small compared to the size of the system. This universal pattern presents an interesting theoretical puzzle: because so few steps separate any pair of nodes, how do these complex systems prevent all information from being everywhere, simultaneously? 

When fitted with the logarithmic scaling function (Table~\ref{tab:table2}), we find that biological networks exhibit the fastest scaling behavior in the network’s mean geodesic distance ($b=1.42\pm0.18$ vs. the next highest parameter $b=0.96\pm0.18$ for informational networks). This scaling behavior of $\langle\ell\rangle$ in biological networks qualitatively connects with its observed scaling of mean degree, i.e.\ they exhibit the weakest positive relationship between mean degree and network size and the strongest positive relationship between mean geodesic distance and network size. This can be intuitively understood as more connected networks generally exhibiting shorter paths between nodes, assuming no strong, systematic constraints on generative processes. The weakest scaling pattern between mean geodesic distance and network size is in technological networks ($b=0.18\pm0.19$). However, as observed with the mean degree, there is substantial variation around the fitted scaling laws in every domain, with the empirical mean geodesic distance sometimes varying by a factor of 2 across networks of a given size $n$.

\paragraph{Clustering Coefficient}
Across all four domains, we observe a generally negative relationship between the clustering coefficient $C$ and network size $n$, as shown in Figure~\ref{fig:empirical_scaling}c, indicating that larger networks tend to exhibit less local transitivity. Notably, Figure~\ref{fig:empirical_scaling}c shows that the scaling behavior of social networks is qualitatively similar to that of the biological and informational domains. While some social networks--- particularly networks from online social media platforms--- maintain clustering coefficients that are approximately an order of magnitude higher than networks of the same size in other domains, there is still substantial structural variation within the domain, contributing to the inter-domain scaling similarity. Triangle densities in technological networks become very small with large $n$, while the other domains maintain non-trivial values.
When fitted with a power law scaling function (Table~\ref{tab:table2}), technological networks exhibit the most pronounced decline in clustering as they grow ($b = -0.5 \pm 0.1$),  while the other three domains decline slower and similarly to each other ($b = -0.3$ for social, biological, and technological domains). Nonetheless, there exists substantial intra-domain heterogeneity: biological networks display structurally distinct clusters, with some small networks exhibiting near-zero clustering and others maintaining relatively high triangle densities. The lower end of clustering in biological systems is often driven by protein interaction networks, which are both sparse and triangle-poor, with two networks in our corpus even exhibiting no triangles at all. These two networks were excluded from the plot in Figure~\ref{fig:empirical_scaling}c. Reflecting the variation in mean degree for networks of the same size, there appears to be a qualitatively separate group of small biological networks, primarily protein interaction networks, with much higher triangle densities, again indicating the existence of structurally different classes of networks within the same domain.

Importantly, the observed inverse scaling of clustering coefficient $C$ with network size (Figure~\ref{fig:empirical_scaling}c) along with the observed logarithmic scaling of mean geodesic distance $\langle\ell\rangle$ (Figure~\ref{fig:empirical_scaling}b) indicates that networks in all four domains exhibit the hallmark properties of small-world networks: short path lengths co-occurring with non-trivial clustering. The broad coexistence of these two properties across networks in all domains indicates that small-world networks (\cite{watts1998collective}) are ubiquitous, and may indicate a universal principle underlying the structure and function of real-world complex networks. That is, the relevant distinction between different domains of networks, e.g., between social and biological networks, is not whether or not they exhibit small-world properties, but rather the degree to which they exhibit them. All domains exhibit the small-world combination of short path lengths and high clustering, but domains vary in the strengths of these two properties.

\paragraph{Degree Assortativity} 
Unlike the other network statistics, degree assortativity $r$ shows little systematic dependence on network size across any of the four domains (Figure~\ref{fig:empirical_scaling}d). On average, networks in our corpus are slightly disassortative, with modest variation across size scales. This suggests that, in contrast to metrics such as mean degree or clustering coefficient, assortative mixing by degree is largely governed by structural factors that are not strongly tied to network scale.

This empirical result stands in contrast to widely cited claims in the literature (\cite{newman2002assortative, newman2003mixing}). Prior studies have suggested that social networks tend to be assortative while technological and biological networks tend to be disassortative, i.e., in social networks, high-degree nodes tend to be connected to other high-degree nodes, while in technological and biological networks they tend to connect to low-degree nodes. Our data reveal no such pattern. Rather, our results show clearly that social networks are not more positively assortative than non-social networks, and instead all four domains exhibit similar scaling behavior in their degree assortativity. In fact, all domains contain networks with positively or negatively assortative networks. For instance, while social networks on average trend positive, our corpus includes several large social systems with neutral or even negative assortativity. Similarly, some biological and technological networks display moderately positive assortativity, contradicting the historical view that disassortative mixing is an intrinsic property of these systems.

When fitted with the logarithmic scaling function (Table~\ref{tab:table2}), informational networks show the strongest---though still modest---positive trend with size ($b = 0.04 \pm 0.03$). Other domains exhibit near-zero slopes: social ($b = 0.02 \pm 0.01$), biological ($b = 0.02 \pm 0.04$), and technological networks ($b = 0.002 \pm 0.03$), respectively. Additionally, we observe that the variance in degree assortativity decreases slightly as network size increases, suggesting a mild convergence toward neutral mixing in larger systems.

Taken together, these findings suggest that degree assortativity is shaped more by network-specific functional or generative constraints than by scale or domain alone. The heterogeneity in observed assortativity values, combined with the lack of strong scaling behavior, underscores the need to reconsider simplistic categorizations of networks based solely on their mixing patterns by degree.

The observed relationships between mean degree and other structural properties---such as mean geodesic distance and clustering coefficient---reveal important qualitative regularities across real-world networks. As shown in Figure 1, networks with higher average degree tend to exhibit both shorter typical path lengths and greater local clustering, while sparser networks are characterized by longer paths and reduced triangle densities. This is particularly striking given that the mean degree, as a first-order statistic, does not capture the full shape of the degree distribution. Nevertheless, it appears to exert a strong influence on more complex features of network structure. These connections underscore the utility of mean degree as a coarse but informative baseline for anticipating other topological properties in empirical networks, and highlight its role in shaping the emergent geometry of real-world systems.

An interesting direction for future work is to investigate the extent to which these empirical scaling laws hold at finer levels of granularity within each scientific domain. To provide an initial assessment, we examined five sub-domains in which our corpus contains at least 20 networks (see Appendix Figure~\ref{fig:appendix_network_diversity}): online and offline social networks within the social domain, protein–protein interaction networks and connectome networks within the biological domain, and software networks within the technological domain. Across these sub-domains, we find that the functional forms of the scaling relationships --- the power-law patterns for mean degree and clustering coefficient, and the logarithmic patterns for mean geodesic distance and degree assortativity --- are broadly consistent with those observed at the domain level. While some quantitative differences do emerge, particularly in the scaling of mean degree, the overall qualitative trends within each sub-domain generally mirror those of their parent domain (see Appendix Figures~\ref{fig:appendix_subdomain_social}, ~\ref{fig:appendix_subdomain_biological}, ~\ref{fig:appendix_subdomain_technological}). These sub-domain analyses therefore suggest that the scaling behaviors we observe at the domain level extend to sub-domains, though substantially larger samples and broader coverage of sub-domains will be needed to evaluate these patterns more fully.

\subsection{Random graph scaling laws}
To what degree can empirical scaling laws in networks be explained by random graph models? We answer this question by (1) fitting a particular type of random graph model to each empirical network, (2) using the fitted models to estimate the expected values for the mean geodesic distance $\langle \ell \rangle$, clustering coefficient $C$, and degree assortativity coefficient $r$, and then (3) fitting a scaling law function to these expected values. Comparing these random graph scaling laws to the corresponding empirical scaling laws allows us to characterize the sufficiency of the underlying random graph assumptions for explaining the observed scaling behavior. Hence, deviations between the random graph and empirical scaling laws helps contextualize the empirical pattern and suggest new insights about how the structure of empirical networks differs from random graphs with particular assumptions.

In each of the three types of random graph models we consider, the mean degree is a specified parameter, and so these models will exactly reproduce their empirical scaling (Figure~\ref{fig:empirical_scaling}a).  On the other hand, the mean geodesic distance, clustering coefficient, and degree assortativity coefficient are not specified parameters, and hence represent useful comparisons.

\subsubsection{Erdős–Rényi random graph}
Across all four scientific domains and all three network measures, the $G(n,m)$ Erdős–Rényi random graph model succeeds in replicating the functional form of all the empirical scaling laws, i.e., its measures scale like logarithms for mean geodesic distance $\langle \ell \rangle$ and degree assortativity $r$, and like a power law for the clustering coefficient $C$. This concordance of functional form suggests that the basic scaling behavior of these network properties can be explained in large part by the randomness of connectivity at the largest scales, since Erdős–Rényi random graphs do not contain structure beyond what is produced by a uniformly random arrangement of edges. At the same time, however, the Erdős–Rényi random graph model scaling does not capture the particular scaling rates that we observe in the data (Table II), indicating that the precise behavior of empirical scaling laws in networks requires more than uniform randomness to explain.

In particular, Figure~\ref{fig:nullmodel_scaling}a shows that specifying the number of edges alone through the Erdős–Rényi $G(n, m)$ model generally results in faster scaling of mean geodesic distance $\langle \ell \rangle$ than is observed in the empirical data. That is, empirical networks tend to be more compact (shorter geodesics) than expected under this random graph model, and this gap widens as networks increase in size. Similarly, real-world networks in every scientific domain have substantially higher clustering coefficients than expected under this model (Figure~\ref{fig:nullmodel_scaling}b). That is, real-world networks are substantially less locally tree-like (more triangles) than Erdős–Rényi random graphs, even as they are also more compact. Finally, across domains, Erdős–Rényi random graphs exhibit degree assortativity values of approximately zero (Figure~\ref{fig:nullmodel_scaling}c), a pattern that diverges slightly from the slight positive logarithmic scaling of degree assortativity in social, biological, and informational networks ($b=0.02$ to $0.04$), but which aligns well with the lack of scaling observed in technological networks ($b=0.002\pm 0.03$). 

\subsubsection{The configuration model}

\begin{figure*}[t!]  
    \centering
    \includegraphics[width=0.99\textwidth]{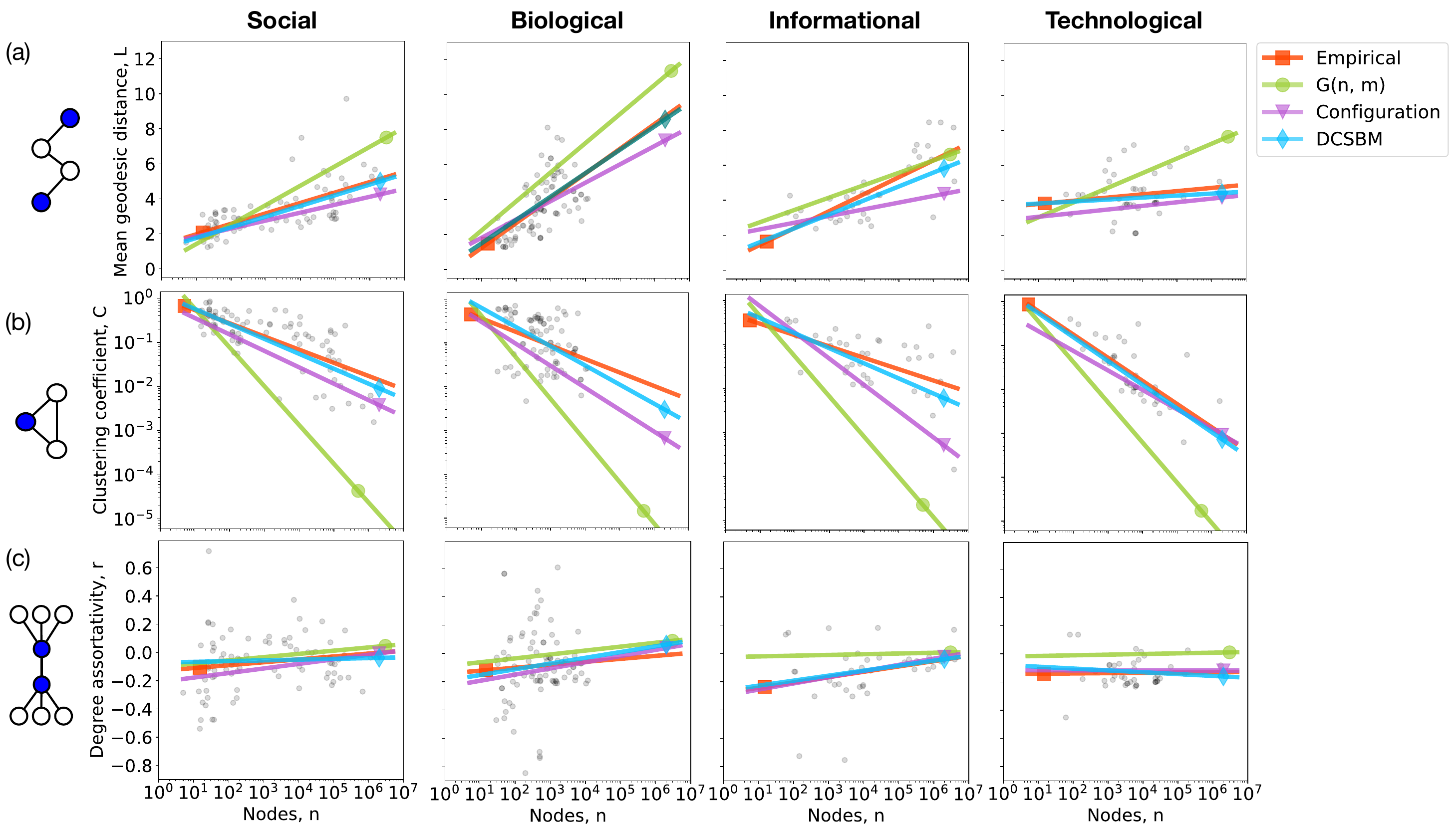}
    \caption{Scaling behaviors of (a) mean geodesic distance $\langle \ell \rangle$, (b) global clustering coefficient $C$, and (c)  degree assortativity $r$, as a function of the number of nodes $n$ for 254 empirical networks. Scaling behaviors are shown for the social, biological, technological, and informational network domains. For a given pair of summary statistic and network domain, the least-squares best-fit line for the empirical networks and for the networks generated from three different random graph models are shown (see legend). For each empirical network, the summary statistics for the random graph models are computed by averaging over 50 random graphs generated from the corresponding model. The scatter points show the empirical networks.}
    \label{fig:nullmodel_scaling}
\end{figure*}

The configuration model generalizes the Erdős–Rényi $G(n, m)$ model by specifying both the number of edges and the degree of each individual node. As such, it is no surprise that the configuration model also replicates the functional forms of the empirical scaling laws: power law for transitivity $C$, and logarithmic for mean geodesic distance $\langle \ell \rangle$ and degree assortativity $r$. However, the configuration model produces scaling laws that are substantially closer in their rates to those of the empirical laws (Figure~\ref{fig:nullmodel_scaling}a-c). That is, the heterogeneity in a network’s degree structure plays an important role in determining the particular rate at which networks change in shape as their size increases.

In all domains, the configuration model slightly underestimates the rate at which geodesic distances tend to grow compared to empirical networks, except perhaps in technological networks (Figure~\ref{fig:nullmodel_scaling}a). That is, empirical networks tend to be slightly less compact than expected under this random graph model, and the gap grows only modestly with network size $n$. This pattern suggests that connections in empirical networks tend to be mildly ``localized'', i.e., they connect pairs of nodes that are otherwise not as distant as a pair of uniformly random stubs in the network, which would tend to lengthen geodesic paths compared to random connections. 

For social, biological, and technological networks, the configuration model comes very close to replicating the rate at which the clustering coefficient scales down with network size, while it still overestimates the rate of decrease for informational networks (Figure~\ref{fig:nullmodel_scaling}b). The similarity of scaling rates between the configuration model and networks in three distinct domains indicates that the observed density of triangles in many empirical networks can be largely explained by those networks’ particular degree structure. This fact is particularly surprising for social networks, which have traditionally been viewed as having far more local clustering (triangles) than can be explained by random graph models (\cite{watts1998collective, newman2003social, bokanyi2023anatomy}). Our results show that this is not exactly the case. Instead, social networks' local clustering falls off about as fast as we would expect for a random graph with the same high mean degree and broad degree distribution. However, the empirical scaling intercept for social networks is still larger than can be explained by the configuration model, meaning that social networks have roughly a constant factor more local clustering than a comparably sized configuration model. Interestingly, biological and informational networks exhibit consistently more transitivity than expected, with gaps that grow with network size. This deviation is not present in social networks, indicating that these networks contain far more local clustering than can be explained by their degree distributions alone.

Finally, in all four domains, the configuration model largely replicates the observed scaling behavior of degree assortativity, being slightly positive in social, biological, and informational networks, and essentially flat in technological networks (Figure~\ref{fig:nullmodel_scaling}c). This close agreement improves over that of the Erdős–Rényi random graph model, indicating that degree assortativity in real-world systems is largely a consequence of their degree structure.

\subsubsection{The Degree-Corrected Stochastic Block Model}

The degree-corrected stochastic block model (DC-SBM) generalizes the configuration model by specifying both the degree sequence of nodes and the community structure of the network. Across domains and network measures, the scaling laws produced by the DC-SBM for our corpus further improve upon those produced by the configuration model and, in several cases, nearly exactly match the empirical scaling laws (Figure~\ref{fig:nullmodel_scaling}a-c). This very close agreement suggests that the degree structure of a network and its particular modular organization together are largely sufficient to explain the particular rates of scaling behavior observed in different kinds of real-world networks.

The DC-SBM almost exactly captures the scaling rates of mean geodesic distances across all domains and all network sizes (Figure~\ref{fig:nullmodel_scaling}a). That is, empirical networks tend to be about as compact as a corresponding DC-SBM network would predict. This correspondence is particularly strong for social and biological networks, while admitting a slight deviation in the scaling rates for informational and technological networks such that empirical networks are slightly less compact (longer geodesic paths) than expected.

The scaling rates for the clustering coefficient under the DC-SBM are substantially closer to those observed empirically than any other random graph model (Figure~\ref{fig:nullmodel_scaling}b). Nevertheless, these scaling laws still predict slightly lower triangle densities than are exhibited by empirical networks as their size increases, with a modest gap appearing for biological networks and a slightly smaller gap appearing for informational networks. In contrast, the DC-SBM provides a very close match to the scaling rates for transitivity in social and technological networks, suggesting that their transitivity rates at any particular network size are largely explained by their degree and community structure.

And, as with the configuration model, the DC-SBM closely replicates the observed scaling behaviors of degree assortativity in all four domains (Figure~\ref{fig:nullmodel_scaling}c). While the configuration model scaling laws underestimated the empirical degree assortativity for smaller social and biological networks, the DC-SBM scaling laws more closely match this behavior, demonstrating a very close agreement with the empirical law across all scales of networks.

The three particular random graph models considered above all have the property that they preserve the exact counts of an empirical network in the synthetic networks we draw from them to estimate the expected values of different network measures. As a robustness check, we conducted the same analyses using modified variants of these models in which those counts are preserved only in expectation. Specifically, we employ the $G(n, p)$ Erdős–Rényi random graph model (\cite{erdds1959random}) which fixes the expected edge density rather than the exact number of edges, the Chung-Lu random graph model (\cite{chung2002connected}) which preserves the expected degree distribution rather than enforcing a strict degree sequence, and the Degree-Corrected Maximum-Entropy Stochastic Block Model which maintains degree sequence and community structure in expectation rather than fixing precise degree sequence and inter-community edge counts (\cite{karrer2011stochastic}). If the agreement between the strict-count random graph models and the empirical scaling laws is due to over-regularizing or over-constraining the space of allowable networks under the models, we should expect to see worse agreement under the expected-count random graph models, which generate a larger range of network structures once estimated.  Despite the relaxed constraints, all of our experimental results are qualitatively the same as those presented here;  see~\ref{sec:appendix_robustness}.

\section{Discussion}

Scaling laws are a useful tool in the study of complex systems: they provide both a means to quantify how network structure or organization systematically varies as the system scales up and a useful empirical reference point for assessing and comparing explanatory models. Using our structurally diverse corpus of real-world networks spanning social, biological, informational, and technological networks, we find robust evidence for four novel empirical scaling laws across networks: (1) the mean degree and (2) transitivity (clustering coefficient) of networks both scale like power laws of the form $y = a\times n^b$, with mean degree increasing slowly ($b>0$) and transitivity decreasing ($b<0$), and (3) the mean geodesic distance and (4) degree assortativity of networks both scale like logarithms of the form $y = a +b\times \log_{10} n$ with $b>0$, except possibly in technological networks, where the degree assortativity appears to be independent of network size.

Across scientific domains, however, we find these patterns exhibit different scaling rates (as measured by the parameter $b$) and intercepts (parameter $a$), suggesting that while the functional forms of the scaling may be universal, domain-specific structural differences determine the particular scaling rates and intercepts within those forms. These findings both affirm and contradict some aspects of conventional wisdom in network science about the structure of real-world networks. First, our findings affirm the belief that real-world networks are almost always highly ``compact,” meaning that pairs of nodes are typically connected by very short geodesic paths, and we show that these short paths have lengths that vary like $O(\log n)$ across networks of different sizes, regardless of what domain the network is from. This empirical fact presents an interesting puzzle: because so few steps separate any pair of nodes, how do these complex systems prevent all information from being everywhere, simultaneously?

In contrast, our findings contradict the conventional wisdom that social networks are fundamentally different from non-social networks in their transitivity (clustering coefficient), i.e., the idea that social networks have high clustering coefficients while non-social networks have very low or nearly zero clustering coefficients. Rather, we show that networks from all domains exhibit the same scaling function---a power law---for how transitivity decreases with network size. This fact implies that social networks are not fundamentally different from non-social networks. However, we do find that the scaling intercept for social networks is substantially higher than for non-social networks. In practice, this difference means that for a given network size $n$, a social network will tend to have about 1.5 times higher transitivity than a biological network of the same size. But, the common scaling behavior means that a biological network of $n$ nodes will have about the same clustering coefficient as a social network with about $4n$ nodes. Hence, while social networks do have more triangles, their local clustering behavior is more a difference in degree, rather than a difference in kind with other types of networks.

Remarkably, the functional forms of the mean geodesic distance and transitivity scaling laws are consistent with those predicted by simple random graph theory, even in social networks. This fact suggests that the randomness of connectivity alone in networks in different domains plays a deep role in shaping their large-scale structure and how it varies with network size. Comparisons with more elaborate random graph models further suggest that the precise scaling rates and intercepts in networks from different domains are largely the result of the particular degree and community (modular) structure of those networks, which itself varies across domains. This finding that the network structure is not totally explained by node degree aligns with the idea of considering network structure at multiple scales (\cite{ugander2013subgraph}) and that the degree sequence alone is a relatively weak constraint on network structure (\cite{alderson2007diversity}). For example, social networks tend to have higher mean degrees than other types of networks, and the social network mean degree itself scales up with network size faster than in other domains. The comparison with more elaborate random graphs also shows that the observed positive scaling of degree assortativity requires these additional structural considerations, as it is not explained by simple random graph theory alone.

The importance of community (modular) structure in explaining the precise scaling rates for empirical networks may suggest an answer to the aforementioned puzzle about what prevents information from being everywhere, all at once: it’s communities. That is, by localizing connections to some degree within communities or modules, networks increase the path lengths only slightly (e.g., compared to the configuration model, which lacks community structure) while providing greater opportunities for information to circulate locally. Developing new mechanistic insights as to how community structure itself emerges in networks remains an under-theorized and under-modeled phenomenon in network science. An alternative possibility is that the reason information is not everywhere all at once is not structural, but rather dynamical. That is, complex systems may have particular, perhaps even domain-specific, dynamical rules that effectively limit the facilitating effects of compact structure on the rapid spread of information. In this way, the possibility of rapid percolation via structural compactness could remain available to the system, even as most of the time, its dynamics make spreading more localized.

These comparisons with random graphs are also suggestive of potential general mechanisms that shape the structure of networks across domains, the identification of which has been a long-running project in network science. However, the fact that the universal functional forms are well-explained by the relatively weak structural assumptions of simple random graph theory suggests that the forms themselves do not require strong mechanistic assumptions to explain. In contrast, the precise scaling behavior within these universal functional forms appears to depend on domain-specific structural mechanisms, although in many cases we find that relatively weak assumptions—such as heterogeneous degree distributions and community structure—are already sufficient to account for most of the variation we observe. Understanding how different specific network formation mechanisms embody the weak structural assumptions we explore here is an important direction for future work, as is evaluating the degree to which existing mechanistic models of network structure agree with these novel empirical scaling laws. Similarly, there remain some interesting deviations between our more elaborate random graph models and the empirical scaling laws, e.g., in the precise scaling rate of transitivity and the mean geodesic distances in informational networks. Investigating other variations of random graph models, such as spatially-embedded growth models (\cite{kaiser2004spatial}), or developing new classes of random graph models that close these gaps, is an interesting direction of future work.

The results presented in our analyses are limited by the diversity and range of the empirical networks that comprise our corpus. While the corpus is relatively large, it has a notable class imbalance, with social and biological networks being overrepresented compared to other domains (Table~\ref{tab:table1}). Furthermore, within the four scientific domains, networks come from different subdomains, e.g., online vs. offline social networks, molecular vs. cellular vs. ecological networks in biology, etc., and the corpus exhibits some subclass imbalance as well (see Appendix Figure~\ref{fig:appendix_network_diversity}). And, the corpus includes no economic or transportation networks, which are two other major domains listed in ICON (\cite{clauset2016colorado}).

The network simplification procedure (\ref{sec:appendix_materials}) discards some information from the original representation of a non-simple network, e.g., edge weights and direction. Graph simplification is a necessary step to enable the comparison of such a wide variety of empirical networks, and the application of a standard and commonly used set of measures of network structure. However, as a result, our interpretation of network properties is sometimes disparate from their original meaning. Prior work (\cite{newman2001scientific}) generally uses context-specific methods to define the concept of a “shortest path” beyond the number of intermediate connections. Expanding the investigation of network scaling laws to include measures that account for edge weights and directions would be an interesting direction for future work.

Our investigation of scaling laws in empirical networks using a large and structurally diverse corpus of real-world networks sheds new light on many old questions in network science, and illustrates the utility of large-scale empirical analyses in identifying “laws” that can guide our intuition about networks and efforts to model them. Our quantitative results lay the groundwork for future work developing more context-specific models of complex systems, grounded by these empirical observations of structure and scaling.

\paragraph{Acknowledgments}
The authors thank Dan Larremore, Brian Keegan, and Cristopher Moore for helpful conversations and guidance. Preliminary versions of these results were presented in part at the International Conference on Network Science (NetSci) in Seoul Korea in 2016, Burlington VT in 2019, and virtually in the SIAM Workshop on Network Science in 2020. We acknowledge the BioFrontiers Computing Core at the University of Colorado Boulder for providing high performance computing resources (NIH 1S10OD012300) supported by BioFrontiers IT.

\paragraph{Funding Statement}
This project was supported in part by Grant No. IIS-1452718 from the National Science Foundation.

\paragraph{Competing Interests}
This work was performed outside of Alexander Ray's current role at Amazon.

\paragraph{Data Availability Statement}
The code supporting this work is available in this \href{https://anonymous.4open.science/r/scaling_laws_repo-4DF1/}{GitHub repository}. The empirical networks used in this work are available \href{https://doi.org/10.5281/zenodo.18676020}{here}.



\paragraph{Author Contributions}
Conceptualization: A.C. Methodology: U.D; A.R; A.C. Data curation: A.R. Experimentation: U.D; A.R. Data visualisation: U.D. Writing: U.D; A.R; A.C. All authors approved the final submitted draft.




\printbibliography
\appendix

\section{Materials}
\label{sec:appendix_materials}

The selection criterion for networks used in this analysis is defined to create a numerically balanced corpus–with regard to relative domain and subdomain proportions– where groups of similar networks do not disproportionately influence end results. Further, as random graphs are used for comparison throughout this analysis, we seek to only include networks without strongly non-random traits.

Networks are filtered from the corpus according to a qualitative understanding of the generative processes that produced them. Specifically, we do not include networks produced by taking a one-mode projection, which induces clique structures, as well as spatially embedded networks, where nodes are embedded in a low-dimensional space, and the existence of edges is constrained by node position. The underlying generative processes of networks in these two categories are such that the probability of a given edge existing is driven by a strongly non-random process, albeit in very different forms. We acknowledge that the vast majority of empirical systems have some non-uniform structure and dynamics, and as such, we rely on qualitative measures to distinguish sufficiently random networks from the rest.

Larger groups of networks that share substantially similar generative processes are not included in their entirety to avoid disproportionate influence on final results. Graph sets related through collection process and context (for example, the Facebook 100 networks (\cite{traud2012social}) or through temporal evolution (sequences of autocorrelated snapshots of a given network) are limited to five arbitrary representatives in the final corpus. Dense networks such as those of Congressional roll-call votes are explicitly excluded from the corpus, as their structure and information are often fundamentally encoded differently than with sparse graphs. We do not restrict our analysis to the largest connected component of each network, ensuring a closer agreement between an empirical graph and the random graph against which we compare it, which is not strictly connected. As the extent to which observed scaling behaviors are explained by degree distribution is of importance in our analysis, the inclusion of the entirety of networks minimizes loss of information.

To ensure a consistent network representation and consistent summary statistics, we ignore edge direction and edge weight and collapse any multi-edges of non-simple networks. We do not include networks with fractional edge weights, as these weights often represent the strength of interaction and thus require a thresholding procedure to simplify. Prior work (\cite{boldi2011hyperanf}) suggests that edge direction should not be ignored in network analysis, in large part because introducing symmetry artificially deflates the mean geodesic distance. The symmetry we introduce to directed networks ensures all components are strongly connected and thus all pairwise paths in a component have a defined distance. Ignoring weights and multi-edges simplifies the analysis, but loses information about the original networks. For example, a network where edge weights correspond to spatial distances and a network where edge weights correspond to the number of interactions between nodes have the same path length interpretation in our analysis, even though their original interpretations are quite different. The simplified path lengths we report are thus only measures of the number of intermediate nodes between endpoints.

\section{Methods}

\subsection{Network Measures}
\label{sec:appendix_network_measures}
We define the mean geodesic distance $\langle \ell \rangle$ of a network as the total sum of all finite pairwise distances, normalized by the total number of reachable vertex pairs, given by:

\[
\langle \ell \rangle = \frac{
  \sum_m \sum_{\substack{ij \in V_m;\ i < j}} d_{ij}
}{
  \sum_m \binom{|V_m|}{2}
},
\]

\noindent where, for each connected component $V_m$ of the network (\cite{newman2010networks}), $d_{ij}$ denotes the shortest-path distance between vertices $i$ and $j$, which is finite only when $i$ and $j$ lie in the same connected component. This definition allows the calculation of $\langle \ell \rangle$ even for disconnected networks.

The global clustering coefficient $C$ is defined as the ratio of triangles to connected triples in the network (\cite{newman2003structure, wasserman1994social}), thus giving the probability that two neighbors of a vertex are connected. The definition of degree assortativity $r$ is that of Equation 4 in \cite{newman2002assortative}.

\begin{figure*}[t!]  
    \centering
    \includegraphics[width=0.99\textwidth]{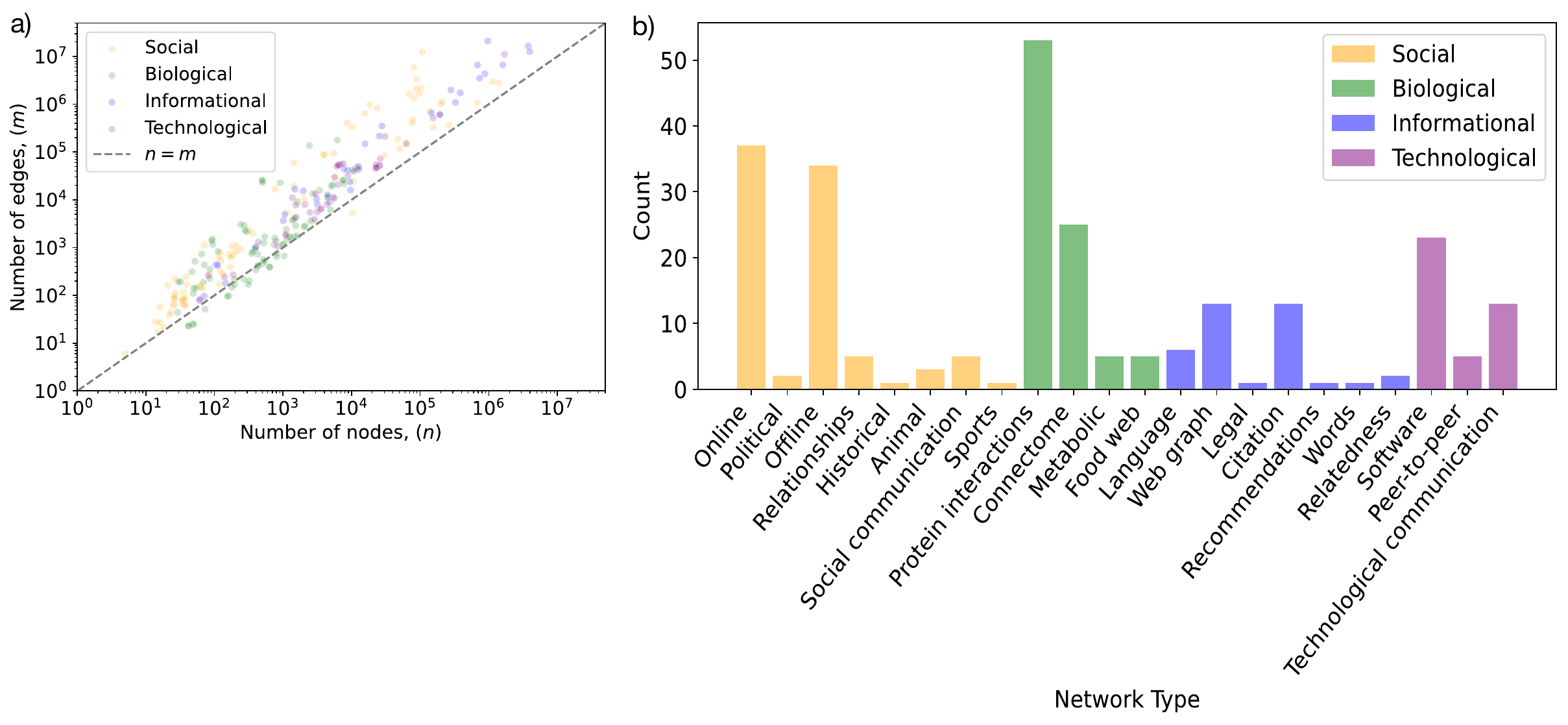}
    \caption{Descriptive overview of the empirical network corpus. (a) Relationship between the number of nodes and number of edges for all networks in the dataset, illustrating the wide range of sizes and densities across the corpus. (b) Distribution of different types of networks within each of the four scientific domains: social, biological, informational, and technological, highlighting the structural diversity of the networks used in our analysis.}
    \label{fig:appendix_network_diversity}
\end{figure*}

The expensive calculation of $\langle \ell \rangle$ for large networks introduces the need for an average path length approximation strategy. While we exactly compute $\langle \ell \rangle$ for all empirical networks in our corpus, the repeated sampling and computation of $\langle \ell \rangle$ for random networks necessitate the use of estimation techniques. Specifically, we estimate the mean of the distance distribution for random networks corresponding to the empirical networks in our corpus. These estimates are then averaged over all samples from the graph space for the null model.

Quickly approximating path lengths in an online fashion is often done with landmark-based methods (\cite{potamias2009fast}). Existing methods for approximating pairwise distance distribution properties–such as the mean pairwise distance– tend to revolve around approximating the neighborhood function of a network. The approximate neighborhood function (ANF) algorithm (\cite{palmer2002anf}) and ANF with HyperLogLog counters (HyperANF) (\cite{boldi2011hyperanf}) utilize probabilistic counting techniques to quickly estimate neighborhood functions; these work extremely well in practice with very large networks (\cite{backstrom2012four}), in large part due to the ease of parallelization and the ability to approximate the neighborhood function with sequential edgelist scans. Other methods include the size-estimation framework (SEF) (\cite{cohen1997size}) and a variety of random sampling techniques (\cite{ye2010distance, crescenzi2011comparison}).

Because the networks we analyze here are small enough to fit in memory, we choose to rely on a simple pairwise distance sampling technique to estimate the mean geodesic distance. Similar sampling methods include the Eppstein-Wang (EW) algorithm (\cite{kosaraju2001proceedings}) adapted to estimating the distance distribution (\cite{crescenzi2011comparison}), which simply uses multiple random breadth-first searches; this method outperforms ANF and SEF for distance distribution approximations and provides absolute error guarantees on the distance distribution approximation through the Hoeffding bound. However, with the EW algorithm, the number of samples needed to achieve a given absolute error is very conservative (\cite{crescenzi2011comparison}) and too slow for our repeated approximations of configuration model random networks. The use of a ``pilot study" to estimate the population standard deviation and thus determine the final sample size is reasonable, though these techniques can yield biased or imprecise estimates (\cite{whitehead2016estimating}).

\begin{figure*}[t!]  
    \centering
    \includegraphics[width=0.49\textwidth]{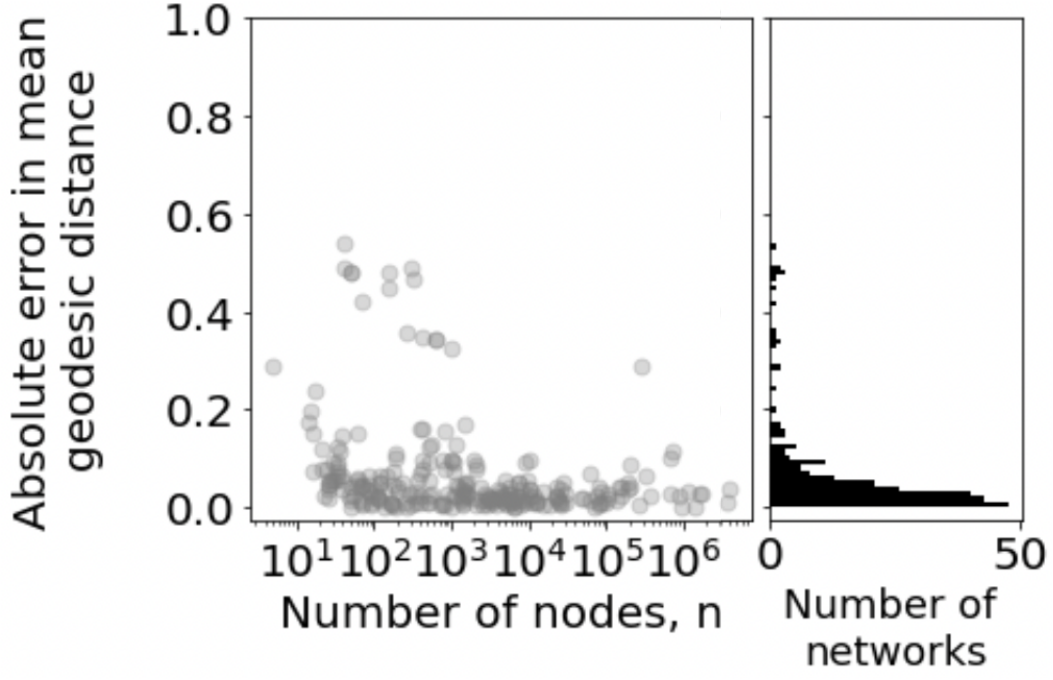}
    \caption{Absolute error, i.e., absolute difference between true and approximated values of mean geodesic distance $\langle \ell \rangle$ using the batch random pairwise distance sampling technique as a function of number of nodes n. Included are 254 empirical networks with exactly computed values of L; approximated values are generated from the batch sampling technique with a batch size of 1000 and a convergence threshold of 0.1. The right panel shows the generated error distribution, indicating that the errors are minimal, and for a vast majority of the networks, this absolute difference is less than 0.2.}
    \label{fig:absolute_error_mgd_with_marginal}
\end{figure*}

Our sampling scheme is similar to the EW algorithm and the random vertex sampling procedure described in \cite{ye2010distance}. We maintain two lists of distances, and sample pairwise distances uniformly at random in batches of 1000, allocating each list half of every batch of samples. We repeat this batch sampling procedure until the means of the two lists fall within a pre-set threshold of each other– we use the value 0.1 throughout this analysis. In practice, this procedure produces very good results (Figure~\ref{fig:absolute_error_mgd_with_marginal}). Accurate estimates of $\langle \ell \rangle$ in a relatively small number of samples may be attributable to the tendency for empirical distance distributions to conform well to the corresponding normal distributions~\cite{ye2010distance}, as well as the fact that we sample from random networks with relatively low average path lengths and correspondingly narrow distance distributions. For many empirical networks, a batch size of 1000 is still quite conservative. We maintain this batch size to balance against the lack of a predetermined sample size; most networks do not require more than one batch to converge, and this ``soft limit" on the number of batches limits the effect of potential multiple testing bias. As Figure~\ref{fig:absolute_error_mgd_with_marginal} illustrates, the magnitude of absolute error is minimal, and for the vast majority of the networks in our corpus, this error is less than 0.2.

\subsection{Null Models}
\label{sec:appendix_null_models}

Each of the three null models used in this analysis--- Erdős–Rényi (\cite{erdHos1960evolution}), configuration (\cite{fosdick2018configuring}), and degree-corrected stochastic block model (\cite{peixoto2017nonparametric})–– implicitly defines a space of networks that share some set of attributes as well as a generative process. To generate the null value of a summary statistic corresponding to a given empirical network and null model, we take the average over 50 networks from the space of graphs defined by that model. The details of the sample procedure used for each model are discussed in detail below. Experimentally, we find there is little variation in the distributions of the averaged summary statistics from random networks in the same graph space; this observation, along with requisite time constraints, influences the choice of 50 samples to construct null values. Note, again, that for all random networks generated in this study, the mean geodesic distance estimation strategy described above is used to estimate $\langle \ell \rangle$.

\paragraph{Erdős–Rényi $G(n, m)$ model} 
\label{para:G_n_m}

We use Erdős–Rényi $G(n, m)$ networks (\cite{erdHos1960evolution}) as a degree-homogeneous null model, which maintains the overall density of the empirical graph by fixing the number of edges in the generated network. To sample from the space of networks with a given empirical density, we simply generate multiple distinct random graphs, with a specified number of nodes $n$ and number of edges $m$.

\paragraph{Erdős–Rényi $G(n, p)$ model}
\label{para:G_n_p}

We also use Erdős–Rényi $G(n,p)$ networks (\cite{erdds1959random}) as a degree-homogeneous null model, in which each of the $\binom{n}{2}$ possible edges is included independently with probability $p$. Unlike the $G(n,m)$ formulation, which fixes the total number of edges, the $G(n,p)$ model fixes the expected edge density of the network. To match the density of an empirical graph with $m$ observed edges, we set $p = \frac{2m}{n(n-1)}$ and generate independent random graphs by sampling each pairwise edge according to this probability.

\paragraph{Configuration model}
\label{para:Config}

We use the well-known configuration model (\cite{fosdick2018configuring}), a degree-heterogeneous null model that preserves the empirical degree sequence exactly. The model generates networks that match the observed degree sequence while randomizing all other structural properties. We sample from the space of simple graphs with a fixed degree sequence using the Markov chain Monte Carlo ``double-edge swap" technique (\cite{dutta2025sampling}). Null values are the average of the given summary statistic over 50 randomly generated networks.

\paragraph{Chung-Lu model}
\label{para:Chung_Lu}

The Chung–Lu model (\cite{chung2002connected}) is a degree-heterogeneous null model that preserves the empirical degree sequence in expectation. Given an observed network with degree sequence ${k_1, \dots, k_n}$, the model defines the probability of an edge between vertices $i$ and $j$ as $p_{ij} = \frac{k_ik_j}{\sum_l k_l}$. Edges are sampled independently according to these probabilities, producing random graphs whose expected degree sequence matches that of the empirical network while randomizing all higher-order structure.

\begin{figure*}[t!]  
    \centering
    \includegraphics[width=0.49\textwidth]{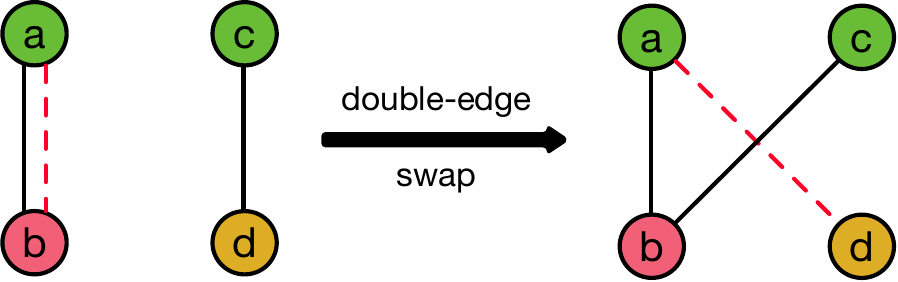}
    \caption{A double-edge swap removing the multi-edge (a, b) by swapping edges (a, b) and (c, d) with (a, d) and (c, b), while still preserving the degree sequence, community labels, and inter-community edge counts. Note that node c is chosen such that nodes a and c belong to the same community.}
    \label{fig:appendix_dcsbm_simplification_heuristic}
\end{figure*}

\paragraph{Degree-corrected stochastic block model (DC-SBM)}
\label{para:DC_SBM_Exact}

We employ the Degree-Corrected Stochastic Block Model (DC-SBM) (\cite{peixoto2017nonparametric}) to generate synthetic networks that preserve both the empirical degree sequence and the block–block edge counts exactly. This process requires estimating three sets of parameters: (i) the degree sequence $k$, (ii) the community labels $B$ assigned to each node, and (iii) the inter-community edge counts $e$. The first step in our approach involves inferring these parameters from empirical networks. While several methods exist for parameter estimation, we adopt a minimum-description-length (MDL) approach (\cite{peixoto2017graph}), which selects the model that minimizes the amount of information required to encode the observed network given the generative model's parameters. Since this inference method is stochastic, it can yield different partitions across runs due to multiple near-optimal solutions having similar description lengths. Given the NP-hard nature of finding the optimal partition, we follow best practices and sample partitions from the posterior distribution rather than selecting a single best-fit partition every time.

To ensure robust inference, we run the MDL-based estimation algorithm 100 times, storing each resulting set of parameters along with its corresponding description length. We then sample 50 parameter sets from the resulting set of parameters with replacement, weighting the selection inversely proportional to the description length—favoring models that provide more parsimonious descriptions of the data. Using the 50 sampled parameter sets, we generate 50 synthetic networks.

It is important to note here that the microcanonical SBM method (\cite{peixoto2017nonparametric}) allows for self-loops and multi-edges in the generated network. Since all our empirical networks are simple graphs in nature, we ensure that the synthetic random graphs are also converted to simple graphs for a fair comparison between empirical and synthetic networks. In order to simplify the synthetic networks, we employ a heuristic, illustrated in Figure~\ref{fig:appendix_dcsbm_simplification_heuristic}, which resolves non-simple edges (self-loops and multi-edges) while maintaining the degree sequence, community labels, as well as inter-community edge counts. Specifically, for each non-simple edge (red dashed line in Figure~\ref{fig:appendix_dcsbm_simplification_heuristic}), we identify a candidate edge swap by selecting nodes $c$ and $d$ uniformly at random such that (i) $a$ and $c$ belong to the same community, and (ii) $c$ and $d$ are connected with an edge. If such nodes exist, we propose to perform a double-edge swap by removing edges $(a, b)$ and $(c, d)$ and replacing them with $(a, d)$ and $(c, b)$. Note that such a rewiring of the edges preserves the degree sequence, community labels, as well as the number of edges in between the communities. However, if edges $(a, d)$ or $(c, b)$ existed before the swap, adding the new edge would result in a multi-edge, which is undesired. In such a case, the swap is rejected, and we reattempt with new candidate nodes. If no valid swap is found after multiple attempts, the non-simple edge is deleted to maintain network simplicity. In our empirical experiments, the proportion of deleted edges was negligible across all network domains: 0.23\% in social networks, 0.37\% in biological networks, 0.2\% in informational networks, and 0.2\% in technological networks. Given that deletions affected less than 0.5\% of edges in all cases, we conclude that this simplification step does not meaningfully alter network structure or impact subsequent analyses.

\paragraph{Degree-corrected Maximum-Entropy Stochastic Block Model}
\label{para:DC_SBM_Expectation}

The degree-corrected maximum-entropy stochastic block model (DCSBM; \cite{karrer2011stochastic}) generalizes the classical stochastic block model by allowing arbitrary degree heterogeneity while preserving community structure in expectation. In this model, each vertex $i$ is assigned a degree parameter $\theta_i$ and a block membership $g_i$, and edges between vertices $i$ and $j$ occur independently with probability $p_{ij} = \theta_i\theta_j\omega_{g_ig_j}$, where $\omega_{rs}$ controls controls the expected density of edges between blocks $r$ and $s$. This model produces random graphs that match both the empirical degree sequence and the community structure in expectation, offering a flexible null model that retains coarse block structure while randomizing all other aspects of the network. The generated random networks are simple networks; hence, network simplification heuristics are not needed.

\subsection{Robustness check of scaling trends with constraint relaxation}
\label{sec:appendix_robustness}

A central feature of the null models used in our main analysis—the Erdős–Rényi $G(n,m)$ model (\cite{erdHos1960evolution}), the configuration model (\cite{fosdick2018configuring}), and the degree-corrected microcanonical stochastic block model (\cite{peixoto2017nonparametric})—is that each reproduces key structural counts of the empirical graph exactly. These strict constraints ensure that every synthetic network sampled from these models matches the observed network in terms of its total number of edges, full degree sequence, or both its degree sequence and block–block edge counts, respectively. While such models provide controlled baselines for comparison, they also severely restrict the space of admissible graphs.

To evaluate whether our empirical scaling relationships depend on this high level of constraint, we repeated our analyses in Figure~\ref{fig:nullmodel_scaling} using more permissive variants of the same families of models—models that impose the same structural targets only in expectation. Specifically, we employed the Erdős–Rényi $G(n,p)$ model (\cite{erdds1959random}), which fixes only the expected edge density; the Chung–Lu model (\cite{chung2002connected}), which matches the expected degree sequence rather than enforcing exact degrees; and the degree-corrected maximum-entropy stochastic block model (\cite{karrer2011stochastic}), which maintains expected degree parameters and expected block connectivity without fixing these quantities exactly. These expectation-based null models generate substantially broader ensembles of graphs and therefore offer a test of whether our results hinge on over-constraining network structure.

\begin{figure*}[t!]  
    \centering
    \includegraphics[width=0.99\textwidth]{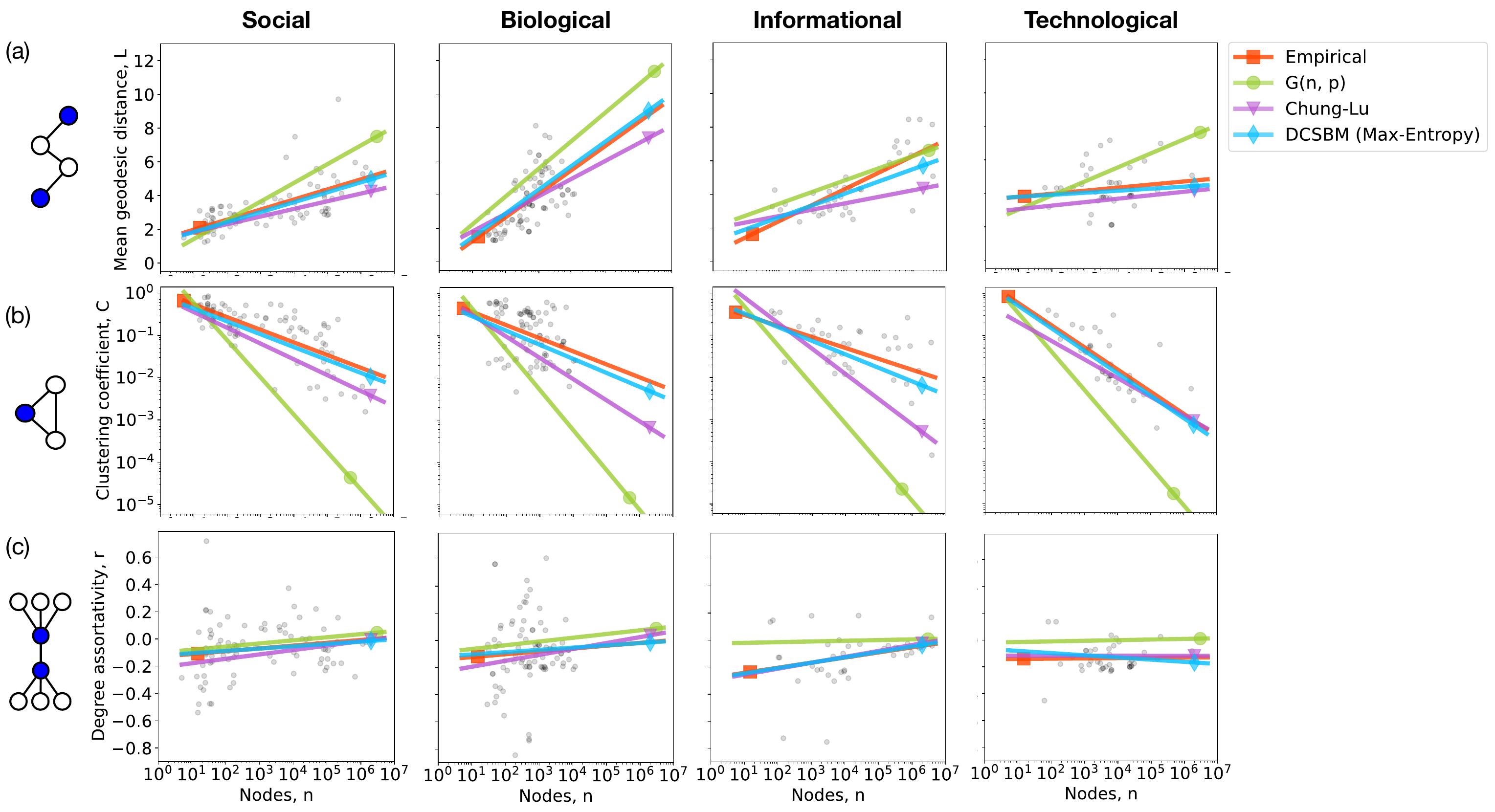}
    \caption{Scaling behaviors of (a) mean geodesic distance $\langle \ell \rangle$, (b) global clustering coefficient $C$, and (c)  degree assortativity $r$, as a function of the number of nodes $n$ for 254 empirical networks. Scaling behaviors are shown for the social, biological, technological, and informational network domains. For a given pair of summary statistic and network domain, the least-squares best-fit line for the empirical networks and for the networks generated from three different random graph models are shown: the Erdős–Rényi $G(n, p)$ model (\cite{erdds1959random}), the Chung-Lu model (\cite{chung2002connected}), and the Degree-Corrected Maximum-Entropy Stochastic Block Model (\cite{karrer2011stochastic}) (see legend). For each empirical network, the summary statistics for the random graph models are computed by averaging over 50 random graphs generated from the corresponding model. The scatter points show the empirical networks.}
    \label{fig:appendix_modelfit_expectation}
\end{figure*}

If the strong agreement observed between the empirical networks and their strictly constrained counterparts were merely an artifact of those tight restrictions, the looser expectation-based ensembles should produce noticeably weaker alignment with the empirical scaling trends. However, as shown in Figure~\ref{fig:appendix_modelfit_expectation}, the qualitative behavior of all scaling relationships remains unchanged. Thus, our findings are robust to substantial relaxation of the underlying null-model constraints, indicating that the observed scaling patterns are not driven by over-regularization but reflect genuine properties of the null models.

\subsection{Scaling trends at sub-domain level}
\label{para:subdomain_trends}

\begin{figure*}[t!]  
    \centering
    \includegraphics[width=0.95\textwidth]{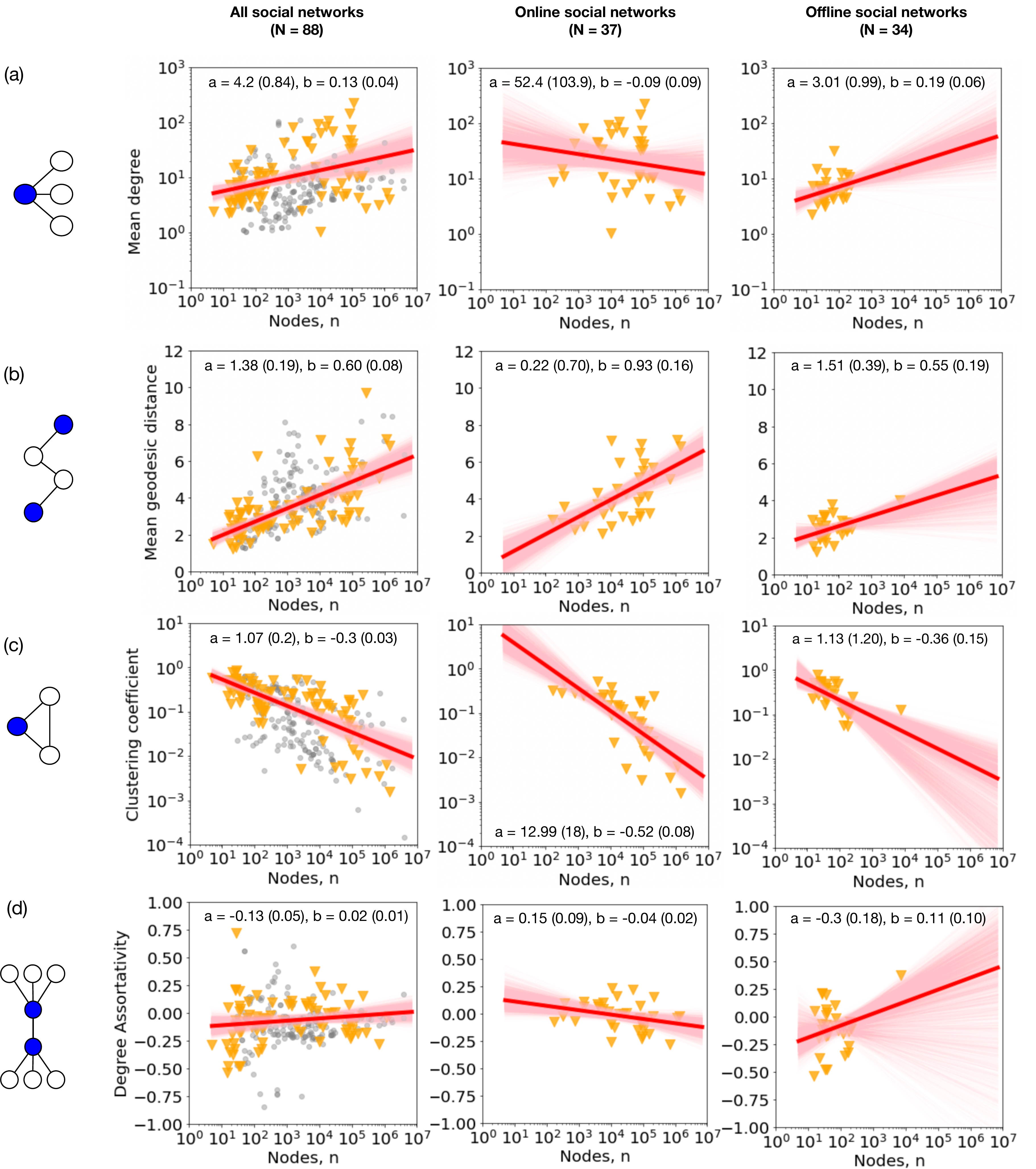}
    \caption{Scaling behaviors of (a) mean degree ($\langle k \rangle = a \times n^b$), (b) mean geodesic distance ($\langle \ell \rangle = a + b \times \log_{10} n$), (c) global clustering coefficient ($C = a \times n^b$), and (d)  degree assortativity ($r = a + b \times \log_{10} n$), as a function of the number of nodes $n$ for empirical social networks. The first column includes all 88 social networks from our corpus, the second includes only the 37 networks belonging to the sub-domain of online social networks, and the last column includes only the 34 offline social networks. Uncertainties correspond to standard deviations of 10000 bootstrap samples of the fitted coefficients.}
    \label{fig:appendix_subdomain_social}
\end{figure*}

\begin{figure*}[t!]  
    \centering
    \includegraphics[width=0.95\textwidth]{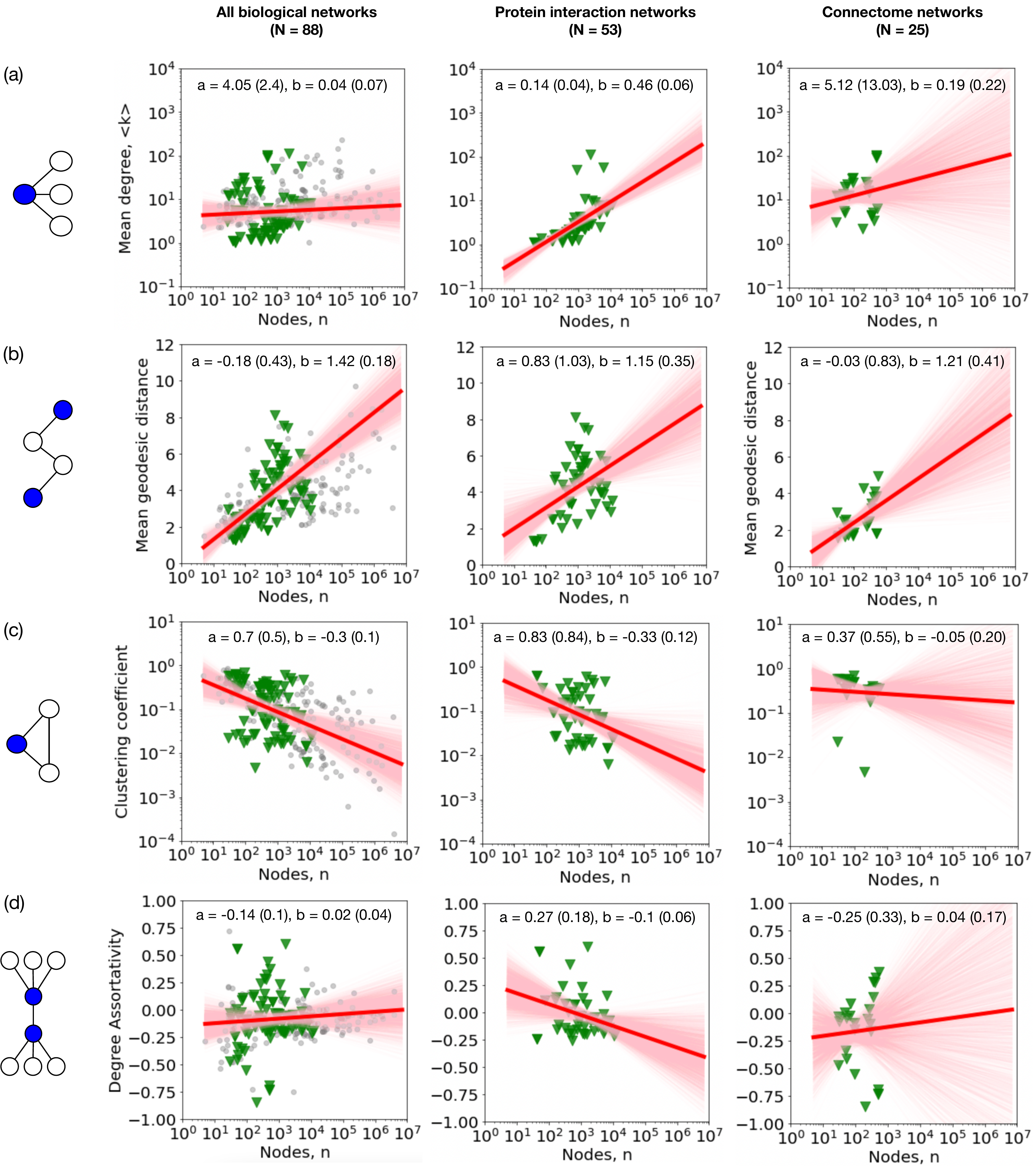}
    \caption{Scaling behaviors of (a) mean degree ($\langle k \rangle = a \times n^b$), (b) mean geodesic distance ($\langle \ell \rangle = a + b \times \log_{10} n$), (c) global clustering coefficient ($C = a \times n^b$), and (d)  degree assortativity ($r = a + b \times \log_{10} n$), as a function of the number of nodes $n$ for empirical biological networks. The first column includes all 88 biological networks from our corpus, the second includes only the 53 protein-protein interaction networks, and the last column includes only the 25 connectome networks. Uncertainties correspond to standard deviations of 10000 bootstrap samples of the fitted coefficients.}
    \label{fig:appendix_subdomain_biological}
\end{figure*}

Figures \ref{fig:appendix_subdomain_social} and \ref{fig:appendix_subdomain_biological} present the scaling behaviors of mean degree, mean geodesic distance, clustering coefficient, and degree assortativity for those social and biological sub-domains for which our corpus has at least 20 networks, namely online and offline social networks, and protein interaction and connectome networks. The qualitative scaling patterns of the sub-domains resemble those at the domain level, though some quantitative deviations are apparent, particularly for the mean degree. These plots provide preliminary support for the robustness of the observed scaling laws across sub-domains. Systematic investigation of a broader range of sub-domains with larger sample sizes remains an important direction of future work for understanding how scaling laws vary across diverse systems.

\begin{figure*}[t!]  
    \centering
    \includegraphics[width=0.7\textwidth]{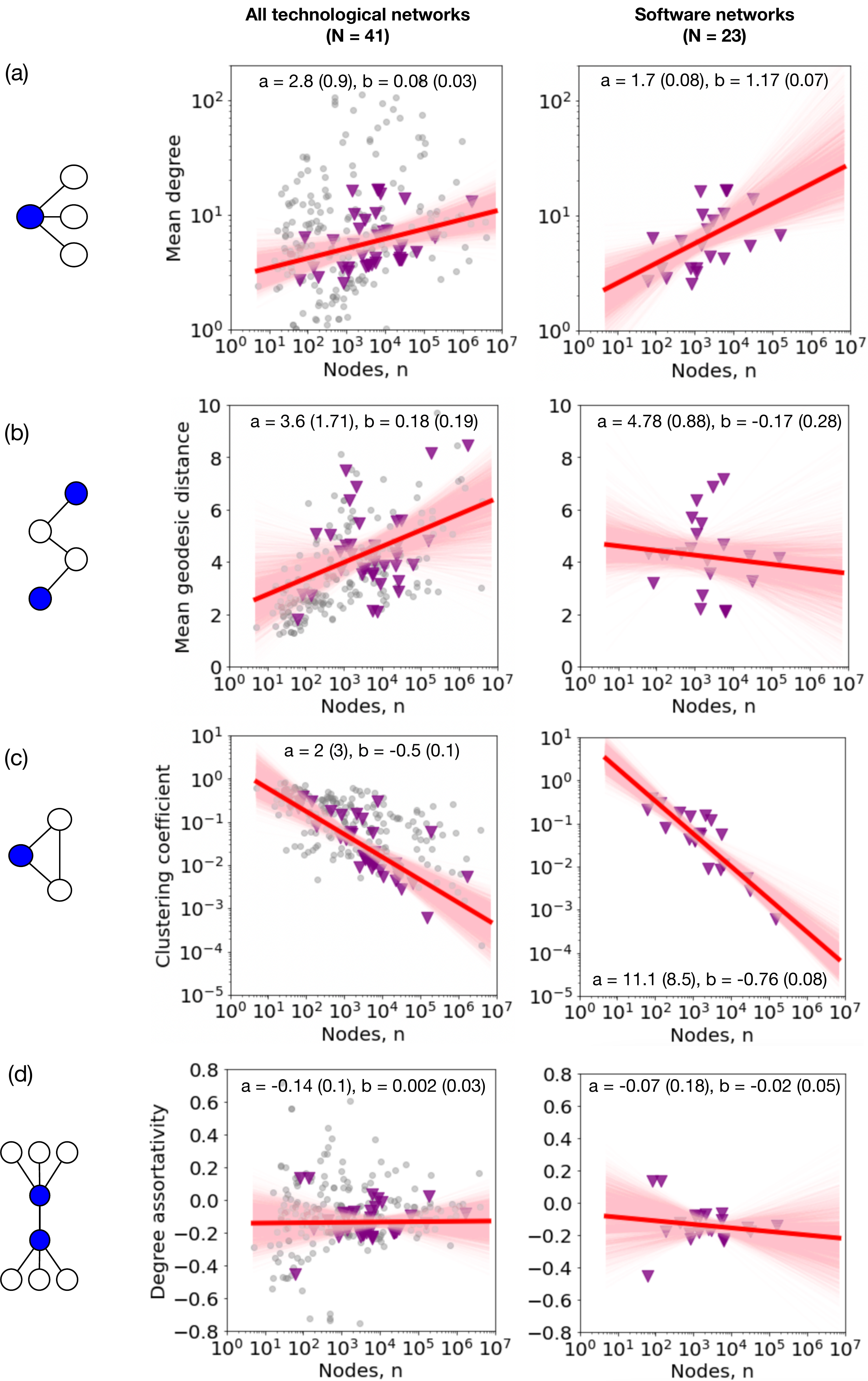}
    \caption{Scaling behaviors of (a) mean degree ($\langle k \rangle = a \times n^b$), (b) mean geodesic distance ($\langle \ell \rangle = a + b \times \log_{10} n$), (c) global clustering coefficient ($C = a \times n^b$), and (d)  degree assortativity ($r = a + b \times \log_{10} n$), as a function of the number of nodes $n$ for empirical technological networks. The first column includes all 41 technological networks from our corpus, and the second includes only the 23 software-specific networks. Uncertainties correspond to standard deviations of 10000 bootstrap samples of the fitted coefficients.}
    \label{fig:appendix_subdomain_technological}
\end{figure*}

\end{document}